# A rock-salt-type Li-based oxide, Li₃Ni₂RuO₆, exhibiting a chaotic ferrimagnetism with cluster spin-glass dynamics and thermally frozen charge carriers

**Sanjay Kumar Upadhyay,[1] Kartik K Iyer,[1] S. Rayaprol[2], P.L. Paulose,[1] and E.V. Sampathkumaran[1,*]**

[1]Tata Institute of Fundamental Research, Homi Bhabha Road, Colaba, Mumbai 400005, India

[2]UGC-DAE Consortium for Scientific Research, Mumbai Centre, R-5 Shed, BARC Campus, Trombay, Mumbai – 400085, India

*Corresponding author: sampath@mailhost.tifr.res.in

**The area of research to discover new Li containing materials and to understand their physical properties has been of constant interest due to applications potential for rechargeable batteries. Here, we present the results of magnetic investigations on a Li compound, Li₃Ni₂RuO₆, which was believed to be a ferrimagnet below 80 K. While our neutron diffraction (ND) and isothermal magnetization (M) data support ferrimagnetism, more detailed magnetic studies establish that this ferrimagnetic phase exhibits some features similar to spin-glasses. In addition, we find another *broad* magnetic anomaly around 40-55 K in magnetic susceptibility ($\chi$), attributable to cluster spin-glass phenomenon. Gradual dominance of *cluster* spin-glass dynamics with a decrease of temperature ($T$) and the *apparent spread* in freezing temperature suggest that the ferrimagnetism of this compound is a chaotic one. The absence of a unique freezing temperature for a crystalline material is interesting. In addition, pyroelectric current ($I_{pyro}$) data reveals a feature in the range 40-50 K, attributable to thermally stimulated depolarization current. We hope this finding motivates future work to explore whether there is any intriguing correlation of such a feature with cluster spin-glass dynamics. We attribute these magnetic and electric dipole anomalies to the crystallographic disorder, intrinsic to this compound.**

The phenomenon of spin-glass ordering in which the magnetic moments are randomly frozen as the temperature is lowered below a characteristic temperature ($T_g$) discovered several decades ago for magnetic impurities in non-magnetic matrices, is commonly observed in many concentrated magnetic systems as well[1,2]. Such a type of magnetic ordering in compounds is usually facilitated by crystallographic order and can also be triggered by geometrical frustration [see, for instance, Refs. 3-5]. Some materials exhibit what has been known as 're-entrant spin-glass behavior'[6,7]; in such materials, the one occurring at a higher temperature is of a ferro/antiferromagnetic type, which can enter into a spin-glass regime with a lowering of temperature with a unique freezing temperature. A few ferromagnetic materials exhibiting spin-glass characteristics have also been labelled 'chaotic' magnetic systems[7]. Evidences for multiple spin-glass transitions are generally scarce barring some exceptions[8-10], and in some systems of this kind[9,10], ferromagnetic clusters behave like spin-glasses. In this article, we provide evidence for an interesting situation in which one sees a gradual dominance of cluster spin-glass features, as though there is no unique freezing temperature, with a decrease of temperature, for a *crystalline* material, viz., Li₃Ni₂RuO₆, which was believed to be a ferrimagnet ($T_C$ =80 K), [Ref. 11]. Our results thus reveal that the ferrimagnetism of this compound is not that simple. This conclusion is based on viewing together the results of ac and dc magnetization and heat-capacity ($C$) as well as neutron diffraction studies. Interestingly, pyroelectric current reported here also exhibits an anomaly, which appears to arise from thermally frozen-in electric dipoles[12-14], in the same $T$-range in which spin-glass-like features appear. But it is not clear to us at present whether these electric dipole and magnetic phenomena are coupled. It may be stated that, following Ref. 11, this compound was not paid much attention in the literature.

The monoclinic crystal structure (space group, C2/c) in which the compound forms is related to that of $Li_2TiO_3$-type rock-salt structure[15]. In this structure, three distinct positions for Li [Li1 (*8f*), Li2 (*4d*), and Li3 (*4e*)] and two different (*4e*) positions for Ti are possible. In the compound under investigation, it was found that Li1 and Li2 positions are occupied by Li and Li3 is occupied by Ni. Ti1 site is occupied by Ni and almost all Ti2 site is occupied by Ru. A fraction (<10%) of Ru and Ni go to Li1 and Li2 sites respectively and Li in turn occupies majority Ni and Ru sites. If one ignores this disorder, the structure essentially consists of $LiO_6$ octahedra alternating with $(Ni_{2/3}Ru_{1/3})O_6$ octahedral planes, running along *c*-direction. It is however clear from the above discussions that there is a significant crystallographic disorder in this material. A view of the crystal structure of this compound, however ignoring disorder, along a-axis is shown in Supplementary Information (see Supplementary Fig. S1 online).

## Results

### Dc magnetization

The results of dc magnetic susceptibility as a function of *T* obtained in a magnetic field of (*H*=) 5 kOe are shown in Fig. 1a. There is a gradual increase of χ with decreasing *T* below 300 K, which is cut off by an upturn below 100 K, which becomes sharper below about 80 K, due to the onset magnetic ordering, followed by a peak around 50 K and finally a fall. Inverse χ exhibits a linear region in a narrow temperature interval (225-300 K), below which there is a deviation from this high-temperature Curie-Weiss behavior, attributable to short-range magnetic correlations; the value of the effective moment obtained from the linear region is about 5.3 μB per formula unit which is very close to that expected (5.57 μB) for high-spin divalent Ni (S= 1) and pentavalent Ru (S= 3/2). The value of the paramagnetic Curie-temperature is found to be about -305 K. These findings are in agreement with those reported by Laha et al[11] by measurements with 1 kOe. However, a further study with low-fields (e.g., *H*= 100 Oe, see Fig. 1b), presented here, for zero-field-cooled (ZFC) and field-cooled (FC) conditions of specimen during measurements offers an insight. While ZFC curve qualitatively resembles that obtained with 5 kOe, the FC curve deviates from this curve below about 100 K, without any downfall even in ZFC curve. χ(FC) continues to increase with a tendency to flatten, only below about 30 K, but not at the onset magnetic ordering. There is a weak peak at about 50 K in FC curve, coinciding with the peak temperature in ZFC curve. It should also be noted that there is a shoulder near 30 K (where FC curve flattens) even in ZFC curve; the ZFC curve additionally shows a shoulder near 80 K, which apparently gets smeared in the FC curve due to the steeper variation in this temperature range. While irreversibility in ZFC-FC curves is a signature of spin-glass freezing, the "delayed" flattening of the FC curve with multiple features as described above already signals complex nature of magnetic ordering.

We have also measured hysteresis loops at low fields (Fig. 1c) and isothermal magnetization up to 140 kOe (Fig. 1d) at selected temperatures. *M(H)* plots continue to increase without any evidence for saturation and thus the ferromagnetic state at high fields could not be obtained. This finding emphasizes that the magnetic ordering can not be of a ferromagnetic-type. The hysteresis loops at 30 and 60 K to show a weak hysteresis, which is a characteristic feature of spin-glasses and ferrimagnetism.

### Neutron diffraction

*Crystallographic Structure*: We have analyzed the crystallographic structure of this compound at room temperature with the ND pattern recorded at 300K. The ND pattern was refined using a structural model given by Laha et al[11]. The observed pattern fits very well to this model. The occupancies for each site obtained from Rietveld refinement is as follows: The Wyckoff site 8f is occupied by Li1 and Ru1 in the ratio 92.5:7.5. There are three 4e sites, two of which are occupied by Li/Ni and Ni respectively, and the third is occupied by Li/Ru. The first 4e site is occupied by Li3 and Ni1 in the ratio 87:13, and the second 4e site is fully occupied by Ni2. The third 4e site is occupied by Ru:Li in 85:15 ratio. The 4d site is occupied by Li:Ni3 in the ratio 87:13. All the oxygen positions are fully occupied. In general, good fits were obtained between calculated and observed ND patterns recorded at different temperatures (300K, 150K, 100K, 65K, 30K, 10K, and 3K).

Magnetic features: In Fig. 2 we have shown the ND data along with Rietveld refinement profile for three selected temperatures, 3, 65 and 150K. The first observation one can make here is that, with decreasing temperature, there is no additional or un-indexed Bragg peak. There is a distinct increase in peak intensity on entering magnetically ordered state (see Supplementary Fig. S2 online). Fig. 3 shows the variation of cell parameters (including the monoclinic angle $\beta$ (in degree)). A decrease in temperature decreases the overall unit cell volume.

Since the neutron diffraction patterns measured down to 3K do not show additional magnetic Bragg peak, the magnetic ordering in this compound could be assumed to be ferrimagnetic with the propagation vector, $\boldsymbol{k}$ = (0 0 0), also taking note of the fact that isothermal magnetization at high fields is not ferromagnetic-like (that is, absence of saturation) as mentioned earlier. Magnetic moments were refined for temperatures well below $T_C$ only. Using BasIreps program of the Fullprof suite[16,17], irreducible representations and basis vectors were obtained for all the magnetic ions at different crystallographic sites. For each magnetic ion, the $\Gamma 1$ representation was sufficient to correctly represent the magnetic structure with reasonable values of the magnetic moment. As per the cationic distribution, Ru1 is found at the site 8f (shared with Li1). However, Ru1 does not seem to possess a magnetic moment and hence was not considered for final refinement. Starting from the ND data measured at 3K, the magnetic moments were refined independently for Ni3 at site 4d, Ni1 and Ni2 at 4e sites and Ru2 at site 4e. This arrangement clearly shows that all moments at site 4e lie on the same plane and the moment on site 4d lie above and below this layer. The coupling of magnetic moments between sites 4d and 4e is anti-parallel, thereby giving rise to ferrimagnetic structure as shown in Fig. 4. The refined magnetic moments for Ni and Ru at different sites are tabulated in Table 1 and shown pictorially in Fig. 5. It can be clearly seen that among the magnetic ions, Ni3 (at 4d site) exhibits negative moments, indicating that these moments are anti-parallel to the rest of the magnetic ions (at site 4e) as clearly seen in the Fig. 5. It therefore appears that it is this antisite Ni which results in net ferrimagnetism. We believe that non-monotonic variation of magnetic moments with temperature could be genuine, considering complex features in the temperature dependence of dc magnetic susceptibility.

## Ac magnetic susceptibility

Fig. 6a shows real ($\chi'$) and imaginary ($\chi''$) parts of ac $\chi$. It is obvious that, following the upturn below 100 K with lowering temperature, there is a peak in both these parts at 82 K for the frequency ($\nu$) = 1 Hz and this peak shifts towards higher $T$ range with increasing $\nu$, for instance by 2 K for 1333 Hz; apart from this, the intensity of the peak also decreases with increasing $\nu$. $\chi''$ also exhibits a $\nu$-dependent peak near 80 K, which is a characteristic feature of spin-glass freezing. This implies that the ferrimagnetism could be a chaotic one, as proposed for another re-entrant ferromagnet long ago[7]. With a further lowering of temperature, a broad peak appears around 53 K in $\chi'$ for $\nu$= 1.3 Hz, which varies with frequency, with this peak-temperature increasing by about 1 K for 1333 Hz. A careful look at the left side of this $\chi'$ peak suggests a weak change of slope near 40 K, as though there is a superposition of at least two peaks below 70 K, as though there is more than one characteristic freezing temperature. In fact, this is more clearly reflected in $\chi''(T)$, which peaks near 40 K for $\nu$= 1.3 Hz. This peak shows an apparent upward shift by a few degrees when measured with 1333 Hz. We also measured ac $\chi$ in the presence of a dc magnetic field (5 kOe) and the above-described features are completely suppressed with a dramatic reduction in the values and with overlapping-curves for different $\nu$ (see Fig. 6a). This is a key support for spin-glass-like dynamics.

## Heat-capacity

In the inset of Fig. 6c, we show the plot of $C(T)$ below 100 K and there is no evidence for any feature down to 1.8 K that can be attributed to long range magnetic ordering. The absence of a feature at the onset of magnetic ordering (near 80 K) may be either due to the fact that rapidly varying large lattice contribution around this temperature obscures the expected λ-anomaly. Crystallographic disorder also can contribute to smearing the feature in the entire temperature range. Therefore, it is difficult to delineate the contributions due to magnetic frustration, though this phenomenon also must play a role for the lack of $C(T)$ anomaly.

**Isothermal remnant magnetization ($M_{IRM}$)**

We measured $M_{IRM}$ at three selected temperatures, 1.8, 30, 65 and 125 K. The specimen was zero-field-cooled to desired temperature, and then a field of 5 kOe was applied. After waiting for some time, the field was switched off, and then $M_{IRM}$ was measured as a function of time ($t$). We find that $M_{IRM}$ drops to negligibly small values within seconds of reducing the field to zero at 125 K; however, it decays slowly with $t$ at other temperatures, as shown in Fig. 6b. These offer support to spin-glass dynamics. The curves could be fitted to a stretched exponential form of the type[18] $M_{IRM}(t) = M_{IRM}(0)[1+A\exp(-t/\tau)^{1-n}]$, where A and n are constants and $\tau$ here is the relaxation time. It is found that the value of n falls in the range $0.5 - 0.7$. The values of relaxation times are rather large (e.g., 100 mins at 2 K and about 28 mins for 30 and 65 K). These values are in fact in agreement with that reported for cluster spin-glasses[18].

**Waiting time dependence of dc magnetization**

We looked for aging effects [see, for instance, Ref. 19] in dc magnetization at two temperatures (30 K and 65 K) characterizing spin-glass phase. For this purpose we have followed ZFC and FC protocols as described, for instance, in Ref. 20. In ZFC protocol, we cooled the sample to the desired temperature, waited for certain period of time, switched on a dc field of 100 Oe and measured the increase of $M$ as a function of time. In FC protocol, the specimen was cooled in 100 Oe, and after waiting for certain period, the decay of $M$ was measured as a function of time after the field was switched off. The curves thus obtained for two waiting times are shown in Fig. 7. It is obvious from this figure that the curves for 3000 s are displaced with respect to those for a lower waiting time of 300 s. This is very distinct at 30 K for both ZFC and FC protocols, establishing spin-glass-like spin dynamics at such low temperatures. For 65 K, this displacement of curves is visible for ZFC protocol, but it is feeble for FC protocol, as though spin-glass-like behavior tends to weaken with increasing temperature in the magnetically ordered state. Clearly aging phenomenon is present in this compound and demonstrates gradual nature of the variations in spin-glass freezing with changing temperature.

**'Memory effect' in dc magnetization**

In order to look for memory effect, we obtained $\chi(T)$ curves in different ways. In addition to ZFC curve in the presence of 100 Oe without a long wait at any temperature (which is a reference curve), we have obtained a ZFC curve after waiting at two temperatures 25 and 60 K for 3 hours each (and also for 6 hours each in another independent experiment). We obtained the difference between these two curves and plotted the same as $\Delta M$ versus $T$ in Fig. 6c. It is distinctly clear that there are clear 'dips' at these two temperatures in this plot. It is found that the intensity of the 'dip' is increased for a wait of 6 hours, with respect to that for 3 hours. This is a signature of frustrated magnetic behavior, even in the ferrimagnetic phase (just below 80 K), as discussed for assemblies of nanoparticles with ferromagnetic core and antiferromagnetic shell[21].

**Complex permittivity and pyroelectric current behavior**

Complex permittivity and pyroelectric current studies also reveal interesting behavior well below $T_C$. Dielectric constants ($\varepsilon'$) and the loss factor ($\tan\delta$) are shown in Fig. 8a below 100 K for two selected frequencies (1 and 100 kHz). Beyond 100 K, $\tan\delta$ increases dramatically and therefore extrinsic contributions tend to dominate. The observation we would like to stress is that both $\varepsilon'$ and $\tan\delta$ undergo a gradual increase with $T$ from 1.8 K, without any apparent peak or any other feature. Therefore, we rule out the presence of any ferroelectricity below 150 K in this compound. We have also measured magnetocapacitance at various temperatures and the changes observed in $\varepsilon'$ for $H= 140$ kOe are 0.01 and 0.1% at 2 and 25 K respectively. Therefore, magnetodielectric coupling is rather weak.

However, $I_{pyro}$ as a function of $T$ exhibits a distinct feature (measured with two poling electric fields 100 and 200 V corresponding to 2.08 kV/cm and 4.16 kV/cm for the sample used). That is, the plot (Fig. 8b) shows a peak at about 40 K for a rate of warming of temperature ($dT/dt$) of 2K/min for the poling by - 2.08 kV/cm at 100 K. The peak gets reversed in sign when poled by +100 V. The intensity of the peak increases for 200 V, as shown in Fig. 8b. This finding mimics that expected for ferroelectricity. However, since we do not find any anomaly in $\varepsilon'(T)$ (Fig. 8a), these peaks can not be attributed to ferroelectricity. To gather further support for this conclusion, we have performed pyroelectric current measurements  for different $dT/dt$. The results (see Fig. 8c) reveal that the peak in fact shifts to higher temperatures with increasing $dT/dt$, for instance, to ~43 K and ~46 K for 5 K/min and 8K/min respectively. Such a strong variation is not expected[13] for ferroelectric transitions. We have also obtained the behavior of $I_{pyro}$ in the presence of a dc magnetic field of 10 kOe and we find (see Fig. 8b) that the intensity of the peak in the plot is dramatically suppressed, as in the case of ac $\chi$.

**Discussion**

From the results presented above, it is clear that there appears to be a contradiction between the conclusion from neutron diffraction results (suggesting well-defined magnetic structure, say, ferrimagnetism) and that from other bulk measurements (in particular,  frequency dependence in  ac susceptibility, and aging, memory and ZFC-FC curves bifurcation behavior in dc magnetization, revealing spin-glass features). Clearly, the magnetism of this compound is very complex. The fact that the peaks in $\chi''$ in the range 40 to 60 K are not cusp-like suggests that there is a spread in the freezing temperatures in this $T$-range.  This spread is consistent with various features noted around 30 - 50 K and 75 K in Fig. 1b. For this reason, it is tempting to claim that this compound could be one of the rare examples for multiple spin-glass freezing phenomenon.  It is not clear whether a relaxation phenomenon of ferrimagnetic structure is operative.

In order to understand the nature of magnetism better, we have  analyzed the ac $\chi$ results in terms of the conventional power law, associated with the critical slowdown of relaxation time, $\tau/\tau_0 = (T_f/T_g - 1)^{-zv}$. Here, $\tau$ represents the observation time ($1/2\pi v$), $\tau_0$ is the microscopic relaxation time, $T_g$ is the spin-glass transition temperature, $T_f$ corresponds to freezing temperature for a given observation time and $zv$ is the critical exponent.  For the feature around 40 K, we  obtained $T_g \approx 28$ K, $zv \approx 7.12$ and $\tau_0 \approx 1.8 \times 10^{-4}$ s. For the one around 80 K, corresponding values are: ~80 K, ~2.6 and ~$1.6 \times 10^{-7}$ s.  For a conventional spin glasses[2], the $zv$ value falls in the range ~4-13, and $\tau_0$ value ranges between $10^{-10}$ and $10^{-13}$ s.  It is clear that the values of $\tau_0$ obtained are in general higher than that for the conventional spin glasses. But the deviation is highly pronounced for the feature around 20-40 K. Judged by these values, one can interpret[22,23] that the ferrimagnetic regions form clusters  exhibiting spin-glass-like inter-cluster dynamics, with the  cluster-glass behavior gradually strengthening with decreasing  temperature.   The parameters derived from isothermal remnant behavior also offers support for  cluster-glass behavior, as described earlier. Therefore, long-range magnetic ordering is labelled  'chaotic'  in this article.

It is interesting to see a feature in pyroelectric current in the glassy magnetic phase. This does not arise from ferroelectricity, as mentioned earlier.  Therefore, an alternate explanation should be offered for the observation of the peak in $I_{pyro}$. At this juncture, it may be recalled that such a dependence of the peak on $dT/dt$ has been explained in terms of 'thermally stimulated depolarization current (TSDC)'[12,14] in the past literature. This phenomenon can be explained as follows: The mobile charge carriers, presumably introduced by crystallographic defects due to intrinsic disorder described above, tend to organise themselves to screen the applied electric field, and, with   a lowering of temperature, these charge carriers get trapped randomly forming electric dipoles and persist for a very long time after removal of the electric field.  These carriers can be released thermally, which appears as a peak in $I_{pyro}$. The type of charge carriers trapped determine the sign of $I_{pyro}$ with respect to that of the electric field. For instance, positive $I_{pyro}$ for a negative electric field implies trapping of negative charges, whereas the same sign for both implies holes-trapping. Thus, in $DyMnO_3$, such a peak was attributed to holes[14], whereas, in the case of yttrium iron garnet[13], electrons are responsible. Therefore, in the present case, considering the opposite sign of the electric field and $I_{pyro}$, one can confidently state that the negative charges get trapped.

It is intriguing to note that the *T*-range over which this phenomenon occurs is essentially the same as that of the spin-glass anomalies in ac χ. The observation that $I_{pyro}$ feature is suppressed by a dc magnetic field of 10 kOe as in the case of ac χ may signal some connection between these magnetic and electric dipole phenomena. However, it is an open question whether this is truly the case or whether it is just accidental. Intrinsic crystallographic disorder must be the root-cause of all these electric and magnetic dipole anomalies.

In short, the present results reveal that the new Li-based compound, $Li_3Ni_2RuO_6$, is not a simple ferrimagnet, but is characterized by growing influence of spin-glass dynamics with decreasing temperature in the magnetically ordered state. Due to strong chemical disorder, intrinsic to this compound, it appears that spin-glass clusters form with apparently different freezing temperatures, rather than a single freezing temperature. Thus, such multiple cluster-glass freezing temperature for a crystalline compound is not commonly reported and we hope this compound would serve as an ideal system to model such a behavior. We find pyroelectric anomalies attributable to thermally stimulated depolarization mechanism, presumably due to trapping of negative charge by crystallographic disorder, though the physical origin for possible connection with cluster-glass dynamics is not obvious at present. Thus this compound exhibits interesting magnetic and pyroelectric anomalies.

**Methods**

Polycrystalline sample was prepared as described by Laha et al[11] by a solid state reaction route. Required amounts of high purity (>99.9%) starting materials, $Li_2CO_3$, Ni oxalate ($NiC_2O_4.2H_2O$), and $RuO_2$, as per the stoichiometry of the compound, were mixed thoroughly, heated at 673 K for 4 h, at 1073 K for 12 h, and subsequently at 1198 K for 12 h with intermediate grindings between these three stages of heating. X-ray diffraction (XRD) pattern (Cu $K_{\alpha}$) confirms single phase nature of the compound. The backscattered electron images of scanning electron microscope (SEM) have been obtained to check the homogeneity of the sample. We have also performed energy dispersive scanning electron microscopic studies to determine the composition, particularly for Ni and Ru, though it is not easy to obtain precise composition for low atomic number elements, Li and O, with the sensitivity of the SEM employed.

*T*-dependent dc magnetization studies were carried out with the help of a commercial superconducting quantum interference magnetometer (Quantum Design, USA) and ac χ study with different frequencies (ν= 1.3, 13, 133, and 1333 Hz) with a ac field of 1 Oe was also carried out with the same magnetometer. Heat-capacity studies were carried with a commercial Physical Properties Measurements System (Quantum Design, USA). The same system was used to measure complex dielectric permittivity using an Agilent E4980A LCR meter with a home-made sample holder with several frequencies (1 kHz to 100 kHz) and with a bias voltage of 5 V; the same sample holder was used for pyroelectric studies with Keithley 6517B electrometer by poling at 100 K with different electric fields. Unlike otherwise stated, all the measurements were performed for zero-field-cooled condition (ZFC from 300 K) of the specimen.

Neutron diffraction measurements were carried out on polycrystalline samples on a focusing crystal based powder diffractometer (PD-3) at Dhruva reactor, Trombay[24]. Sample was filled in a vanadium can subjected to temperature variation using a closed cycle refrigerator. Neutrons at a wavelength of 1.48Å were used for the diffraction experiments. ND patterns were analyzed for nuclear (crystalline) and magnetic structures by the Rietveld refinement method using the Fullprof program[16,17,24]
.

Table 1:     The values of magnetic moments of Ni and Ru atoms at 4d and 4e sites at three temperatures are tabulated.  Ni1 and Ni2 are at **4e**: (0 y ¼) with different y-positions and Ni3 is at **4d**: (¼ ¼ ½). Ru is also found at site 4e, with different y position. The variation in the y-position with respect to temperature is also shown in the table along with the magnetic moments of Ni and Ru at these different sites.

| Temperature (K) | Ni1@4e | Ni2@4e | Ni3@4d | Ru2@4e |
|---|---|---|---|---|
| 3 | *(y = 0.0913)* 0.925 | *(y = 0.4230)* 0.849 | *(¼ ¼ ½)* -0.485 | *(y = 0.7423)* 0.601 |
| 10 | *(y = 0.0906)* 1.22 | *(y = 0.4243)* 1.439 | *(¼ ¼ ½)* -0.131 | *(y = 0.7422)* 0.821 |
| 30 | *(y = 0.0910)* 1.11 | *(y = 0.4223)* 1.22 | *(¼ ¼ ½)* -0.283 | *(y = 0.7412)* 0.653 |
| 65 | *(y = 0.0962)* 1.024 | *(y = 0.4199)* 1.047 | *(¼ ¼ ½)* -0.137 | *(y = 0.7442)* 1.199 |

**Author contributions**

**Competing financial interests**

The authors declare no competing financial interests.

**Figure 1 | Magnetization data for $Li_3Ni_2RuO_6$.** Dc magnetic susceptibility as a function of temperature measured in a magnetic field of (a) 5 kOe, and (b) 100 Oe are plotted in (a) and (b) respectively. In (a), inverse susceptibility is also plotted with a line through the Curie-Weiss region. In (b) the curves obtained for ZFC and FC conditions are shown Low-field hysteresis loops at 30 and 60 K and isothermal magnetization extended to high fields at 1.8, 30 and 60 K are also shown in (c) and (d) respectively.

**Figure 2 | Neutron diffraction patterns of $Li_3Ni_2RuO_6$ measured at 150, 65 and 3K.** The data for 3 and 65K include a fitting for the magnetic structure, as described in the text.

**Figure 3 | Temperature dependence of different unit cell parameters obtained from the Rietveld refinement of neutron diffraction patterns is plotted.** Lines are drawn through the data points as a guide to the eyes.

**Figure 4 | The magnetic structure of $Li_3Ni_2RuO_6$ at 3K.** The arrow in red colour represents magnetic moment of Ni3 (at 4d site) ions, whereas arrows in blue and cyan represent magnetic moments of Ni1 and Ni2 (both at 4e site) respectively. The Ru2 (at 4e site) moment is shown in green colour.

**Figure 5 |** The values magnetic moments at different sites in $Li_3Ni_2RuO_6$ structure are plotted as a function of temperature.

**Figure 6 | (a) Real and imaginary parts of ac susceptibility measured with various frequencies (1.3, 13, 133 and 1333 Hz), (b) isothermal remnant magnetization at 1.8, 30 and 65 K, and (c) the difference in magnetization curves, ΔM, obtained with and without waiting at 25 and 60 K for $Li_3Ni_2RuO_6$.** In (a), the arrows show the direction in which the peaks shift with increasing frequency and with the omission of data points through the lines, In the inset of (c), heat-capacity as a function of temperature is shown. The $\chi''$ curves in (a) for 133 and 1333 Hz are shifted along y-axis (by 0.01 and 0.02 emu/mol respectively), for the sake of clarity.

**Figure 7 | Dc magnetization as a function of time** (waiting time dependence or aging experiments) for zero-field-cooled (ZFC) and field-cooled (FC) protocols as described in the text for $Li_3Ni_2RuO_6$ for 30 and 65 K.

**Figure 8 |** Temperature dependence of (a) dielectric constant ($\varepsilon'$) and loss factor ($\tan\delta$) shown for two frequencies (1 and 100 kHz), (b) pyroelectric current, $I_{pyro}$, for two poling fields, obtained by increasing the temperature at the rate of 2K/min, and (c) $I_{pyro}$ as a function of $T$ for different rates of change of $T$, after poling with 2.08 V/cm, for $Li_3Ni_2RuO_6$. In (b), the curve obtained in a field of 10 kOe is also included.

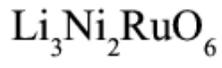

Li$_3$Ni$_2$RuO$_6$

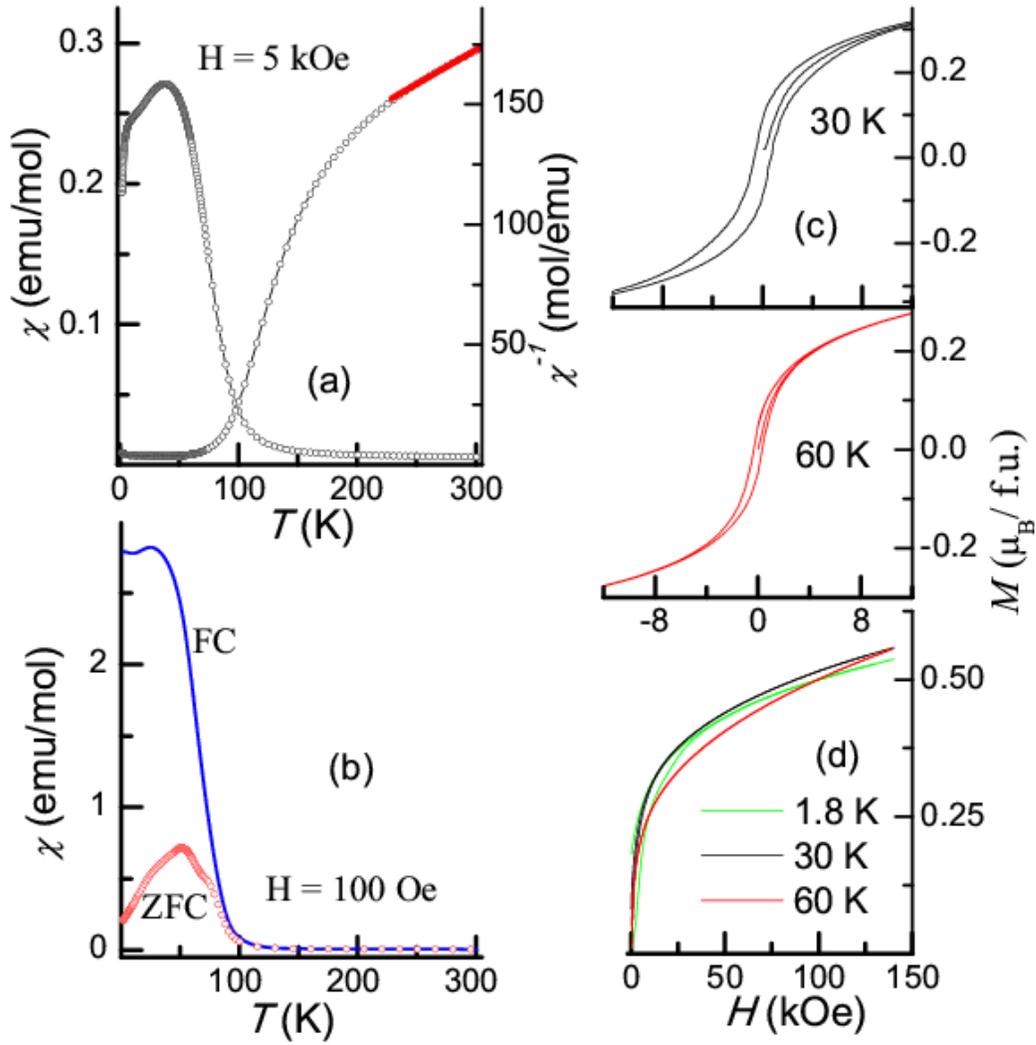

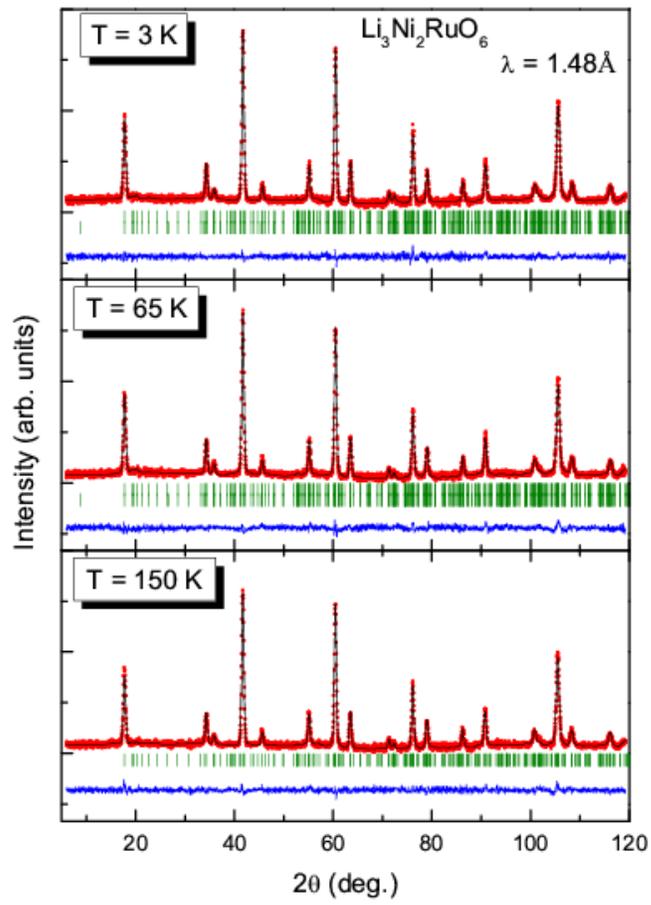

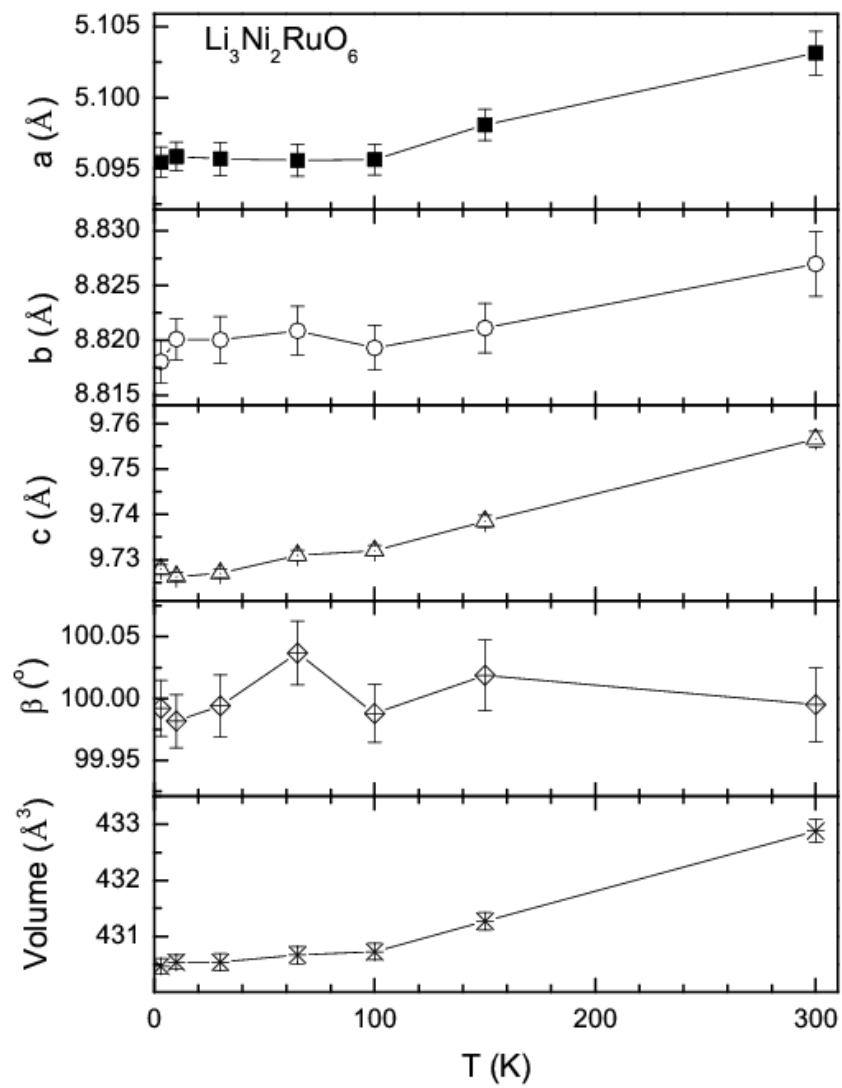

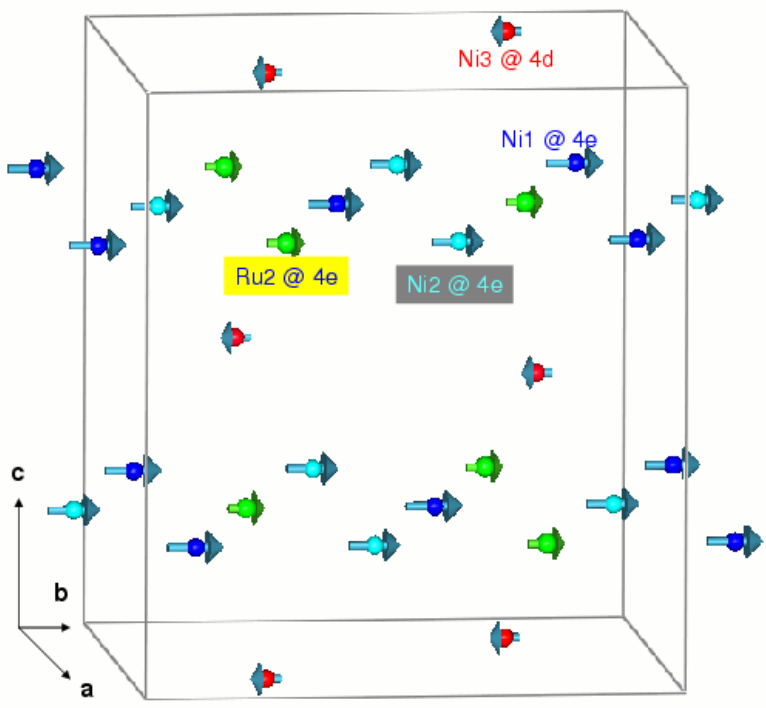

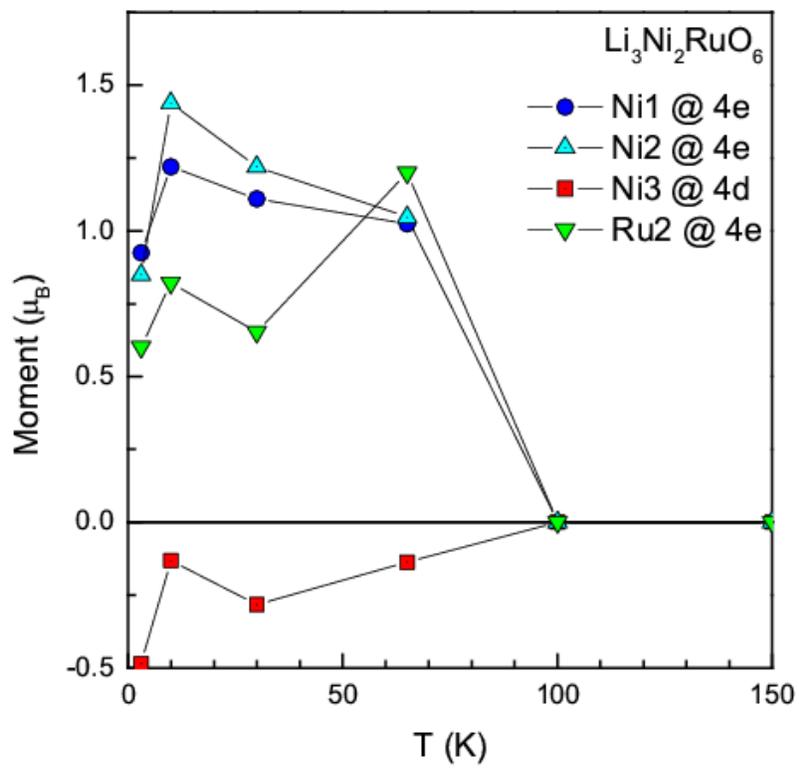

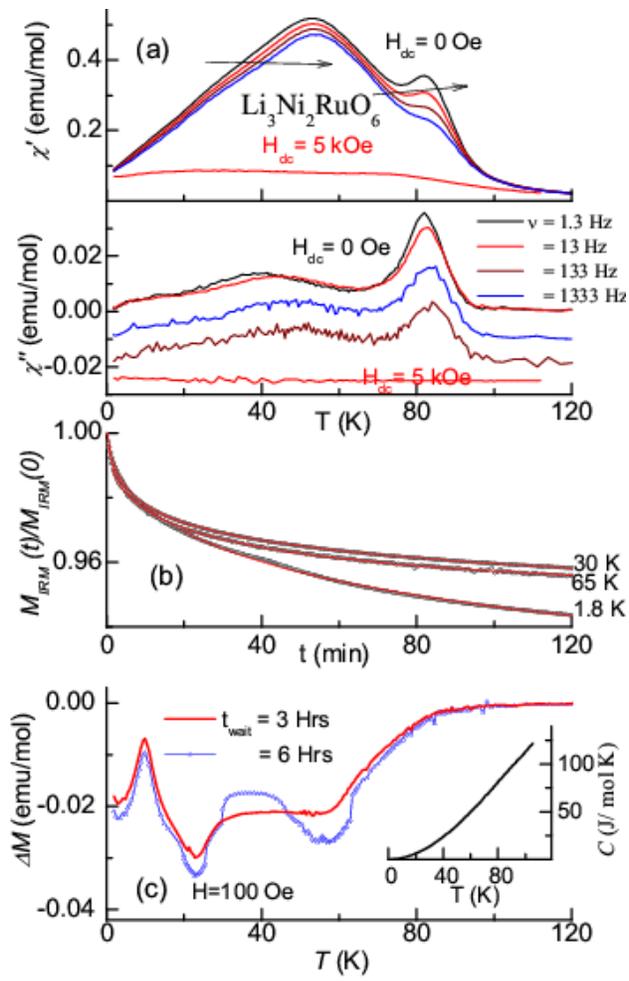

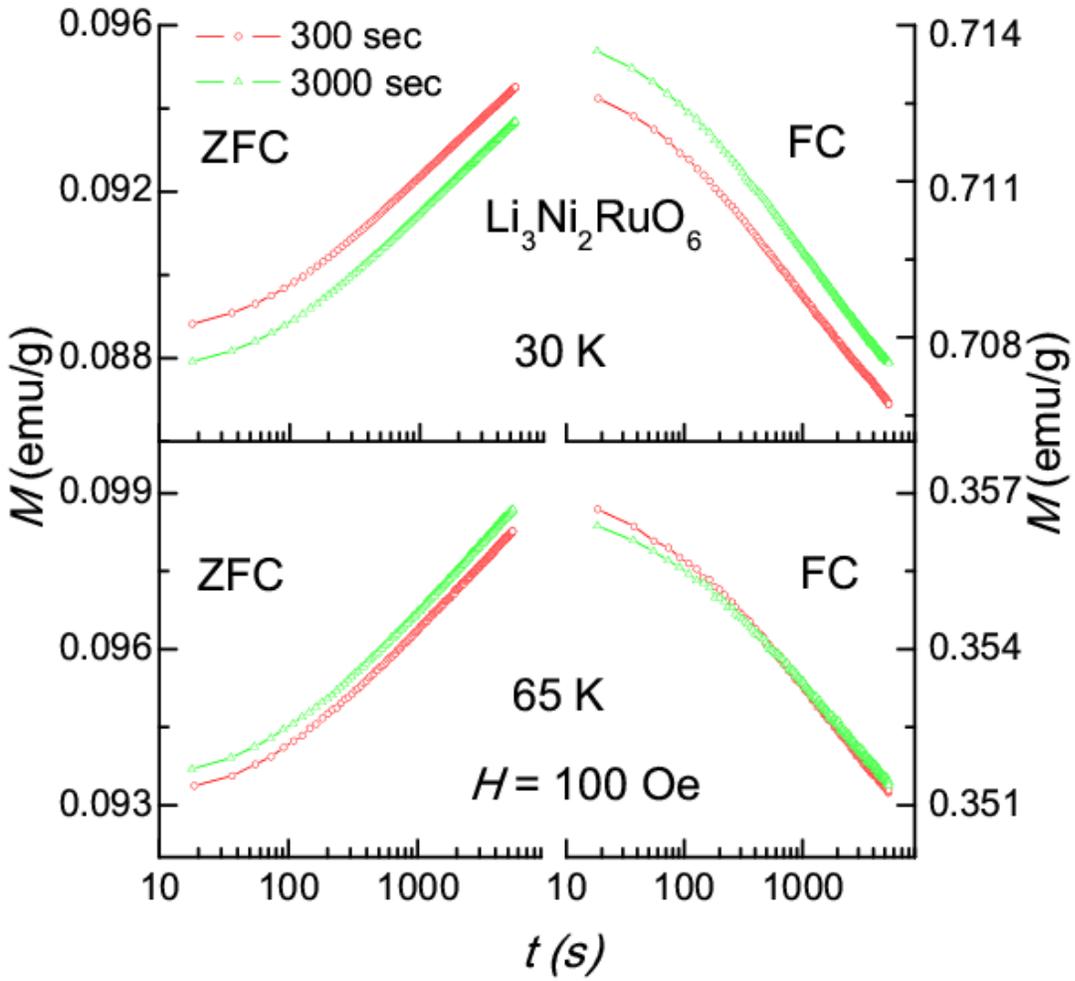

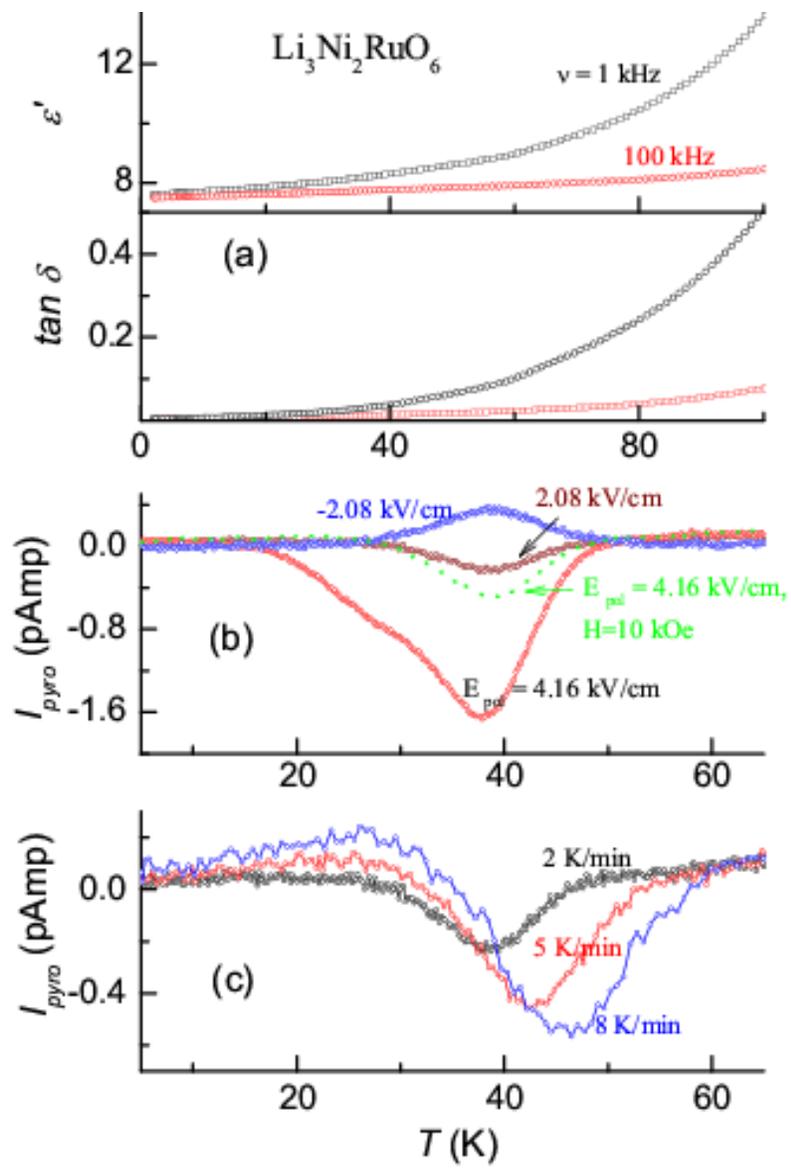

**Supplementary Information**

**A rock-salt-type Li-based oxide, Li₃Ni₂RuO₆, exhibiting a chaotic ferrimagnetism with cluster spin-glass dynamics and thermally frozen charge carriers**

**Sanjay Kumar Upadhyay,[1] Kartik K Iyer,[1] S. Rayaprol[2], P.L. Paulose,[1] and E.V. Sampathkumaran[1,*]**

[1]Tata Institute of Fundamental Research, Homi Bhabha Road, Colaba, Mumbai 400005, India

[2]UGC-DAE Consortium for Scientific Research, Mumbai Centre, R-5 Shed, BARC Campus, Trombay, Mumbai – 400085, India

*Corresponding author: sampath@mailhost.tifr.res.in

Here, we show crystal structure of Li₃Ni₂RuO₆ along the [100] direction (Fig. S1) and also compare raw neutron diffraction data for three different ranges in an expanded scale (Fig. S2) at three temperatures. In Fig. S1, we ignored crystallographic disorder discussed in the article, for the sake of simplicity. Fig. S2 clearly brings out subtle variations in the intensity.

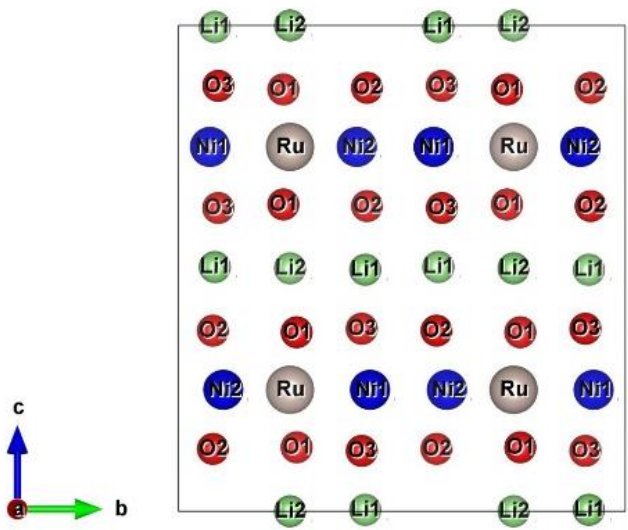

**Supplementary Figure S1:** Crystal structure of Li₃Ni₂RuO₆ viewed along the [100] direction.

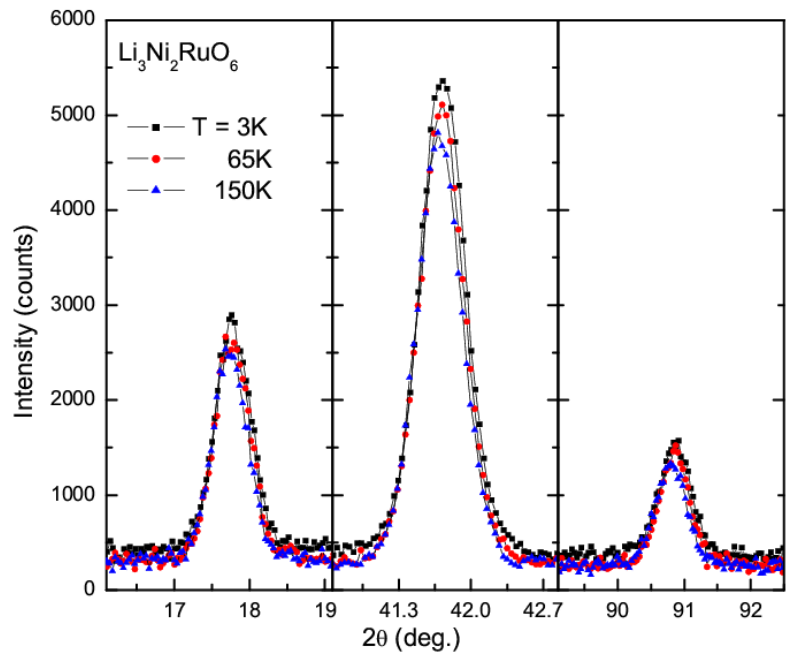

**Supplementary Figure S2:** The raw data of the neutron diffraction patterns recorded at T = 3, 65 and 150K are plotted on the same scale for three different regions.